 \documentclass[12pt]{article}
 \usepackage{graphicx}
 \usepackage{amssymb}
 \usepackage{amsmath}
 \usepackage{amsfonts}
 \usepackage{revsymb}
 \newcommand{\be}{\begin{equation}}
\newcommand{\ee}{\end{equation}}
\newcommand{\bea}{\begin{eqnarray}}
\newcommand{\eea}{\end{eqnarray}}
\newcommand{\bean}{\begin{eqnarray*}}
\newcommand{\eean}{\end{eqnarray*}}

\begin{document}
\begin{titlepage}
\bigskip
\rightline{}

\bigskip\bigskip\bigskip\bigskip
\centerline {\Large \bf {Dynamics of First Order Transitions with Gravity Duals}}
\bigskip\bigskip
\bigskip\bigskip

\centerline{\large  Gary T. Horowitz and Matthew M. Roberts}
\bigskip\bigskip
\centerline{\em Department of Physics, UCSB, Santa Barbara, CA 93106}
\centerline{\em  gary@physics.ucsb.edu, matt@physics.ucsb.edu}
\bigskip\bigskip
\begin{abstract}
A first order phase transition usually proceeds by nucleating bubbles of the new phase which then rapidly expand. In confining gauge theories with a gravity dual, the deconfined phase is often described by a black hole. If one starts in this phase and lowers the temperature, the usual description of how the phase transition proceeds violates the area theorem. We study the dynamics of this phase transition using the insights from the dual gravitational description, and resolve this apparent contradiction.
\end{abstract}
\end{titlepage}
\baselineskip=16pt
\setcounter{equation}{0}
 \section{Introduction}
 
Confining - deconfining transitions in certain gauge theories have been given  geometric descriptions using the gauge-gravity correspondence \cite{witten-thermal}. A common feature of these descriptions is that the deconfined phase is described by a black hole, and the confined phase is described by a nonsingular spacetime without a horizon. In this paper, we study the dynamics of this transition in the case when it is first order. One might expect that if one starts at high temperature in the deconfined phase, and then lowers the temperature below its critical value, one would nucleate bubbles of confined phase which grow until they collide. On the gravity side this would correspond to starting with a black brane, nucleating ``gaps" in the horizon, which then grow until they take over the entire horizon. This scenario has been implicitly assumed in some papers in the literature \cite{Creminelli:2001th,Kaplan:2006yi}. 
 
This scenario, however,  has an serious problem: it violates a well established result in gravitational physics, namely the area theorem.
The original nucleation of the bubbles of confined phase decreases the horizon area, but one could argue that this is a quantum process, so the area theorem does not apply. The more serious problem is with the growth of the bubbles, which is usually assumed to be rapid and modeled on Coleman's discussion of false vacuum decay \cite{Coleman:1977py}, or its generalization to nonzero temperature \cite{Steinhardt:1981ct} . However, at large $N$, the growth of the bubbles should be described by a classical supergravity solution which obeys the area theorem.

We will show that if one starts with a supercooled deconfined phase (with temperature well below the critical temperature), the bubbles of confined phase can grow rapidly for a while, during which time the horizon increases its extent in the fifth direction to compensate for the decrease in area in the field theory directions.  However, when the temperature of the black hole reaches the critical temperature, the growth slows dramatically. The result is a localized black hole which is dual to a plasma-ball \cite{aharony-pballs}.  Plasma balls are localized regions of deconfined  plasma at (or near) the critical temperature surrounded by the confining vacuum. In the large $N$ limit, they are stable. When $1/N^2$ corrections are included, plasma-balls decay slowly by thermally radiating hadrons at the deconfinement temperature. This is dual to Hawking radiation of the localized black hole in the bulk.\footnote{Previous discussions of the confining-deconfining transition as a ``slow" process focussed on the rate of bubble nucleation \cite{Cline:2000xn,Randall:2006py}. We are discussing a qualitatively different effect which occurs after the bubbles are present.}

By now there are several examples of confining gauge theories that have a purely gravitational dual description \cite{witten-thermal,Klebanov:2000hb,Maldacena:2000yy}. We will focus on perhaps the simplest example which consists of ${\cal N} = 4$ super Yang-Mills compactified on a  circle with antiperiodic fermions. This breaks the supersymmetry, and gives mass to the fermions and scalars. The low energy limit is  a confining, purely bosonic 2+1 dimensional gauge theory.
We will show that in this case, the bubbles of confined phase must stop expanding rapidly when the black hole still occupies at least one quarter of the initial volume! At that point, the evolution depends on the boundary conditions. If the system is strictly held at a constant temperature (below the critical temperature), the black hole will slowly evaporate, so the phase transition is eventually completed over a long time scale. However, since AdS acts like a confining box, it seems more natural to work in a microcanonical ensemble and not to include an external heat bath. In this case, we will see that the black hole cannot evaporate completely unless $N^2 $ is sufficiently small. Since the semi-classical gravitational description is only valid for much larger values of $N^2$, the gauge theory will always contain a region of deconfined phase.

There is a related question which we do not address here arising in
recent attempts to understand RHIC physics  using gauge-gravity duality.  The RHIC fireball is often modeled by a black hole (see e.g. \cite{Nastase:2006eb} and references therein) , but as it expands and cools it rapidly enters the confining phase.  How can the horizon disappear so quickly? Lattice calculations indicate that ordinary QCD does not have a first order deconfinement transition, so this is not a problem of bubble nucleation. Nevertheless, it is hard to imagine a consistent gravity description in which a macroscopic event horizon  is present  one moment and gone soon after.

\setcounter{equation}{0}  
 \section{Classical gravity analysis}
 
As mentioned above, we will focus on the case of  ${\cal N} = 4$ super Yang-Mills on ${\bf R}^{3}\times S_\theta^1$. The circle has antiperiodic fermions, but is spacelike. It does not represent a euclidean time direction. If the length of this circle is $L$, the (confining) ground state of this theory is described by the AdS soliton \cite{witten-thermal, energyconj}
 \be 
 ds^2=\frac{r^2}{\ell^2}( -dt^2+dx^2+dy^2+fd\theta^2)+\frac{\ell^2}{r^2}f^{-1}dr^2+\ell^2d\Omega_5
 \ee
 where
 \be\label{deff}
 f=1-\frac{r_0^4}{r^4}
 \ee
The space only exists for $r > r_0$ and regularity at $r=r_0$ requires $r_0=\pi\ell^2/L$. At sufficiently high temperatures $T > T_c$, the system is in a  deconfined phase  described by the black 3-brane metric
\be ds^2=\frac{r^2}{\ell^2}(-f dt^2+dx^2+dy^2+d\theta^2)+\frac{\ell^2}{r^2}f^{-1}dr^2+\ell^2d\Omega_5\ee
where $f$ is again given by (\ref{deff}) but now $r_0=\pi\ell^2T$. 

The confinement-deconfinement transition temperature can be obtained by comparing the free energy of a gas at temperature $T$ in the soliton background with the black brane.  To regularize the infinite volume of the spacelike directions, we will compactify $x$ and $y$ onto a (comparatively large) torus, i.e. ${\rm Vol}(T^2)=V_2$, $\sqrt{V_2}\gg\ell,L$. To leading order in a saddlepoint approximation, the free energy is proportional to the action of the euclidean solutions. 
It is clear that the euclidean versions of these geometries are the same, and there is simply a choice of labeling, i.e., which of the two boundary circles is the euclidean time circle $\tau$ with length $\beta = 1/T$ and which is our original spacelike circle $\theta$. As such, we expect a phase transition between the black brane and  thermal gas at $\beta=L$. In fact, we can calculate the free energy difference between the phases using $F\propto S_\mathrm{Euclidean}$ and find:
 \be F_\mathrm{brane}-F_\mathrm{soliton}\propto 1-(LT)^4 \ee
 So for $T<1/L$, the thermal gas dominates, and at $T>1/L$, the black brane  dominates. In this gravitational language, this is usually called the Hawking-Page phase transition \cite{Hawking:1982dh}.
 
Suppose we start in the deconfined phase of the gauge theory and quickly lower the temperature below $T_c = 1/L$.  How does the phase transition proceed?\footnote{Of course a phase transition can not truly occur (at finite $N$) unless we take the volume to infinity, but this is just $V_2\rightarrow\infty$.} We expect bubbles of  confined phase to be nucleated and grow.  On the gravity side this corresponds to bubbles of soliton being nucleated on the black 3-brane. At large $N$, the growth of these bubbles are described by supergravity. We will first consider the growth of these bubbles at fixed energy (i.e. in a microcanonical ensemble) which is the natural boundary condition for asymptotically AdS solutions. At the end, we will comment on the difference between this and evolution in a canonical ensemble.
  
The energy and entropy of the black brane are \cite{Gubser:1996de}
 
 \be E_\mathrm{brane}=\frac{3}{8}\pi^2N^2LT^4V_2\label{braneen}\ee
 \be S_\mathrm{brane}=\frac{A_\mathrm{horizon}}{4}=\frac{1}{2}\pi^2N^2LT^3V_2\label{braneent}\ee
 and the energy of the AdS soliton is \cite{energyconj}
 \be E_\mathrm{soliton}=-\frac{1}{8}\frac{\pi^2 N^2}{L^3}V_2\ee
which of course has no horizon and thus no instrinsic entropy. These energies are measured relative to  pure AdS. Note that there is a mass gap between the soliton and even a low temperature black brane.

After the bubbles are nucleated, only a fraction $\alpha$ of the initial  volume $V_2$ in the $(x,y)$ directions will be occupied by the black brane and the rest will be the soliton.  While it is difficult to follow the subsequent evolution in detail, we know that in the classical (large $N$) limit, the horizon area must increase. So we will fix the energy
\be\label{defE}
 E=\alpha E_\mathrm{brane}+(1-\alpha)E_\mathrm{soliton}
 \ee
 and maximize the entropy with respect to $\alpha$. 

We ignore effects coming from the boundary between these two regions. This transition region is expected to have a scale no larger than $L$ (set by the critical temperature), so it is negligible whenever both regions are much larger than $L$. When the bubbles are first nucleated their size is indeed of order $L$, but the boundary effects quickly become negligible when the bubbles expand. Since $V_2 \gg L^2$,  a bubble initially takes up only a small fraction of the volume of the brane.

The entropy of our configuration is just $\alpha$ times the entropy of the homogeneous black brane:
\be S=\frac{1}{2}\pi^2N^2LT^3\alpha V_2\ee 
where $T$ is fixed by the fact that the total energy $E$ is fixed. It is easier to characterize  $E$ by the equivalent temperature, $T_0$, that a black brane covering all of $V_2$ would have with this total energy. Note that $T_0$ is only well-defined if $E > 0$ so we would have enough energy to form a homogeneous black brane. Since we are starting with a black brane with small regions  of soliton this is indeed the case. Our energy constraint  (\ref{defE}) is then
\be\frac{3}{8}\pi^2 N^2 LT_0^4 V_2=\frac{3}{8}\pi^2 N^2 LT^4\alpha V_2-\frac{1}{8}\frac{\pi^2N^2}{L^3}(1-\alpha) V_2 \ee
This yields
\be\label{solT} 
T=\frac{1}{L}\left[\frac{1-\alpha}{3\alpha}+\frac{1}{\alpha}(T_0L)^4 \right]^{1/4}\ee
and hence
\be S=\frac{\pi^2N^2V_2}{2L^2}\alpha\left[ \frac{1-\alpha}{3\alpha}+\frac{1}{\alpha}(T_0L)^4\right]^{3/4}\ee
Since $\alpha \in [0,1]$, we see two distinct phases: for $T_0\ge 1/L$, the entropy is maximized by the brane covering the entire $V_2$. However since we are starting with a black brane below the critical temperature with part of its volume replaced by the soliton, the physically interesting regime is $T_0< 1/L$. In this case, $S$ is maximized at
\be\label{solalpha}
\alpha=\frac{1}{4}[1+3(T_0L)^4]
\ee
Most notably we have even as $T_0 \rightarrow 0$,   $\alpha=1/4$, so the maximal entropy configuration is a localized black hole taking up a quarter of the initial volume. Therefore even if we start at an energy well below the critical energy, the black brane will \emph{not} disappear completely, but instead will localize. In terms of the gauge theory, there will still be a large region of deconfined plasma. 

Substituting (\ref{solalpha}) into (\ref{solT}) we see that in the maximal entropy configuration, the temperature of the black hole is
\be\label{solTequil}
 T=\frac{1}{L}.\ee
The maximum entropy configuration always has the black hole at the critical temperature, independent of $T_0$! This agrees with the discussion of plasma-balls in \cite{aharony-pballs} where it is shown that the thermodynamic pressure (driving the plasma ball to expand or contract) vanishes at the critical temperature.

\begin{figure}[htp]
\centering
\includegraphics{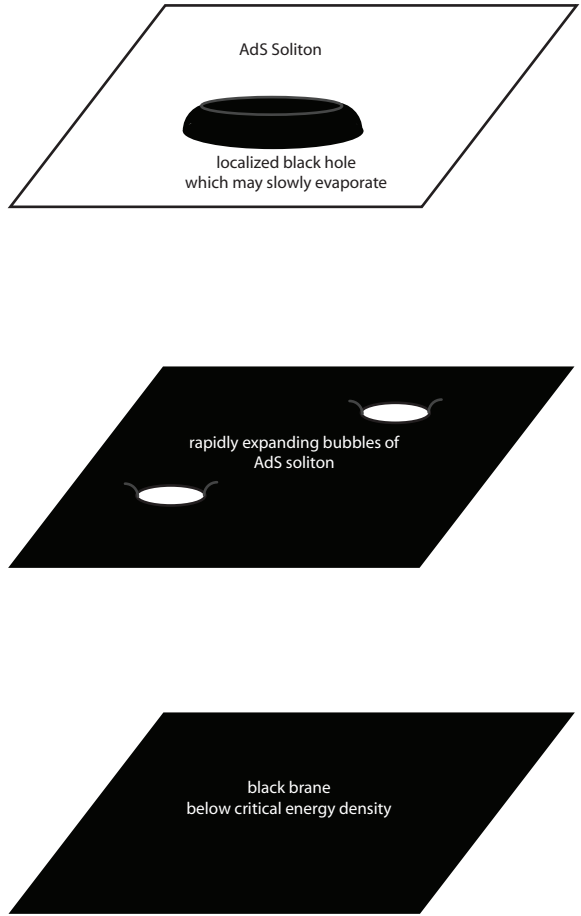}
\caption{The phase transition between a black brane everywhere and a localized black hole.}\label{fig:phasetr}
\end{figure}

\setcounter{equation}{0}
\section{$1/N^2$ effects of a thermal gas}

Since our discussion so far has used just classical gravity, it does not include the entropy in the Hawking radiation around the black hole, or the entropy of the gas in the AdS soliton. We now consider these $1/N^2$ effects. 

A full calculation of the energy and entropy of a thermal gas of the ``massless" degrees of freedom is daunting and not terribly enlightening. We know that a massless scalar field in the soliton background has modes in the AdS radial direction which simply gives us a  tower of particles with masses of order $1/L$ \cite{Constable:1999gb}.  A proper calculation in the black brane phase would require computing the expectation value of the quantum stress energy tensor in the Hartle-Hawking state.

However, the qualitative effects of including Hawking radiation are simple to understand. The large localized black holes have negative specific heat, just like  small Schwarzschild-like black holes \cite{aharony-pballs}. When we include $1/N^2$ effects, the localized black hole  will radiate a small amount of energy and come into thermal equilibrium with its Hawking radiation. It does not completely evaporate. Its  mass  will be slightly reduced, and its temperature will be slightly increased. 

We have focussed on evolution at fixed energy corresponding to a microcanonical description. What would change if we considered a canonical ensemble at fixed temperature? Of course, in this case,  the entropy of our system can decrease (provided the free energy also decreases, so the entropy in the heat bath grows to compensate it).  The initial evolution of the bubbles will be similar. However the localized black hole occupying at least one quarter the volume will no longer be stable. If there is an external heat bath below the transition temperature, the black hole will slowly evaporate. The phase transition will eventually be completed, but over a much longer time scale.

\setcounter{equation}{0}
\section{Discussion}

We have discussed the dynamics of first order confining-deconfining transitions in gauge theories with gravity duals. Starting with a supercooled deconfined phase, there are three stages in the transition (see Fig. 1): (1) nucleation of bubbles of confined phase, (2) rapid growth of these bubbles, (3) large plasma-ball phase. In a microcanonical (fixed energy) description, the plasma-ball is stable and the phase transition is not completed. In a canonical (fixed temperature) description, the plasma-ball slowly hadronizes. In either case, when the system enters stage three, the plasma ball occupies at least 25\% of the initial volume\footnote{This is for the specific case we considered. We expect other systems will be similar.}, so it is not a small effect. 

Starting in the confined phase and raising the temperature, the phase transition  proceeds differently. Now there is no analog of the third stage. Once the temperature is above the critical temperature, the black hole horizon can grow rapidly until it covers the entire volume.

If the rate of bubble nucleation is very slow, and one starts with a black brane well below the critical temperature, it might Hawking evaporate to the point where the size of the circle at the horizon reaches the string scale. It was argued in \cite{Horowitz:2005vp} that a winding string tachyon instability would then set in and cause the black brane to turn into the AdS soliton. We now check that the total entropy is increased in this transition. If we assume that the energy all goes into a gas of particles, then it is easy to see that the entropy would {\it not} increase when $N$ is large. For a black brane with $T \ll 1/L$, the energy released in the transition to the soliton is essentially independent of $T$ and given by the mass of the soliton relative to AdS:  $ E \sim N^2 V_2/L^3 $. Let us focus  on the $N$ dependence. A relativistic gas in $9+1$ dimensions has $S_{\rm gas} \sim E^{9/10} $, so $S_{\rm gas} \sim N^{9/5}$. Since the entropy of the black brane is proportional to $N^2$, $S_{\rm gas} < S_{\rm brane}$ for large $N$, even when the tachyon instability sets in. Another problem with the energy going into a gas is that the effective temperature of that gas would be much greater than the critical temperature. Since the number of species in the bulk is independent of $N$, $E \sim T^{10}$ implies $T \sim N^{1/5}$. With such a high temperature, the gas should collapse back to a black brane. 

The resolution is that the energy released in the transition to the soliton goes mainly into one (or more) highly excited fundamental strings. A highly excited string has entropy $S_{\rm string} \sim E_{\rm string}\  l_s$. If the string is localized near the tip of the soliton,\footnote{This restriction only affects an overall coefficient in the entropy.} its energy is redshifted by a factor $r_0/\ell = \pi \ell/L$. So the proper energy of the string is larger than the change in the total energy measured at infinity:
\be
E_{\rm string} = \frac{\pi N^2 V_2}{8 L^2 \ell}
\ee
The entropy in the string is then 
\be 
S_{\rm string} \sim \frac{N^2 V_2}{L^2} \frac{l_s}{ \ell}
\ee
We want to compare this with the entropy of the black brane at the point where the tachyon instability sets in. The radius of the circle at the horizon is $L r_0/\ell = \pi \ell (LT)$. When this is equal to the string scale, 
\be
S_{\rm brane} \sim \frac{N^2 V_2}{ L^2} \left (\frac{l_s}{ \ell}\right )^3
\ee
Since $\ell \gg l_s$, the entropy in the fundamental string is always greater than the black brane when the winding tachyon condenses. However both of these entropies are much smaller than the localized black hole we discussed earlier. If the bubble nucleation rate is not negligible, the endstate will be the localized black hole with entropy 
\be
S_{\rm local\ bh} \sim \frac{N^2 V_2}{L^2}
\ee

\vskip 1cm
\centerline{\bf Acknowledgements}
\vskip .5cm
It is a pleasure to thank E. Silverstein for discussions. This work was supported in part by NSF grant PHY-0555669.


\begin{thebibliography}{99}

\bibitem{witten-thermal}
  E.~Witten,
  ``Anti-de Sitter space, thermal phase transition, and confinement in  gauge
  theories,''
  Adv.\ Theor.\ Math.\ Phys.\  {\bf 2}, 505 (1998)
  [arXiv:hep-th/9803131].
  %%CITATION = HEP-TH 9803131;%%
  
  %\cite{Creminelli:2001th}
\bibitem{Creminelli:2001th}
  P.~Creminelli, A.~Nicolis and R.~Rattazzi,
  ``Holography and the electroweak phase transition,''
  JHEP {\bf 0203}, 051 (2002)
  [arXiv:hep-th/0107141].
  %%CITATION = HEP-TH 0107141;%%
  
  %\cite{Kaplan:2006yi}
\bibitem{Kaplan:2006yi}
  J.~Kaplan, P.~C.~Schuster and N.~Toro,
  ``Avoiding an empty universe in RS I models and large-N gauge theories,''
  arXiv:hep-ph/0609012.
  %%CITATION = HEP-PH 0609012;%%
  
  %\cite{Coleman:1977py}
\bibitem{Coleman:1977py}
  S.~R.~Coleman,
  ``The Fate Of The False Vacuum. 1. Semiclassical Theory,''
  Phys.\ Rev.\ D {\bf 15}, 2929 (1977)
  [Erratum-ibid.\ D {\bf 16}, 1248 (1977)].
  %%CITATION = PHRVA,D15,2929;%%
  
  %\cite{Steinhardt:1981ct}
\bibitem{Steinhardt:1981ct}
  P.~J.~Steinhardt,
  ``Relativistic Detonation Waves And Bubble Growth In False Vacuum Decay,''
  Phys.\ Rev.\ D {\bf 25}, 2074 (1982).
  %%CITATION = PHRVA,D25,2074;%%

\bibitem{aharony-pballs}
  O.~Aharony, S.~Minwalla and T.~Wiseman,
  ``Plasma-balls in large N gauge theories and localized black holes,''
  Class.\ Quant.\ Grav.\  {\bf 23}, 2171 (2006)
  [arXiv:hep-th/0507219].
  %%CITATION = HEP-TH 0507219;%%
  
  %\cite{Klebanov:2000hb}
\bibitem{Klebanov:2000hb}
  I.~R.~Klebanov and M.~J.~Strassler,
  ``Supergravity and a confining gauge theory: Duality cascades and
  chiSB-resolution of naked singularities,''
  JHEP {\bf 0008}, 052 (2000)
  [arXiv:hep-th/0007191].
  %%CITATION = HEP-TH 0007191;%%
  
  %\cite{Maldacena:2000yy}
\bibitem{Maldacena:2000yy}
  J.~M.~Maldacena and C.~Nunez,
  ``Towards the large N limit of pure N = 1 super Yang Mills,''
  Phys.\ Rev.\ Lett.\  {\bf 86}, 588 (2001)
  [arXiv:hep-th/0008001].
  %%CITATION = HEP-TH 0008001;%%
  
  %\cite{Nastase:2006eb}
\bibitem{Nastase:2006eb}
  H.~Nastase,
  ``More on the RHIC fireball and dual black holes,''
  arXiv:hep-th/0603176.
  %%CITATION = HEP-TH 0603176;%%

\bibitem{Cline:2000xn}
  J.~M.~Cline and H.~Firouzjahi,
  ``Brane-world cosmology of modulus stabilization with a bulk scalar field,''
  Phys.\ Rev.\ D {\bf 64}, 023505 (2001)
  [arXiv:hep-ph/0005235].
  %%CITATION = HEP-PH 0005235;%%

%\cite{Randall:2006py}
\bibitem{Randall:2006py}
  L.~Randall and G.~Servant,
  ``Gravitational waves from warped spacetime,''
  arXiv:hep-ph/0607158.
  %%CITATION = HEP-PH 0607158;%%
 
  \bibitem{energyconj}
  G.~T.~Horowitz and R.~C.~Myers,
  ``The AdS/CFT correspondence and a new positive energy conjecture for
  general relativity,''
  Phys.\ Rev.\ D {\bf 59}, 026005 (1999)
  [arXiv:hep-th/9808079].
  %%CITATION = HEP-TH 9808079;%%
 
 
 %\cite{Hawking:1982dh}
\bibitem{Hawking:1982dh}
  S.~W.~Hawking and D.~N.~Page,
  ``Thermodynamics Of Black Holes In Anti-De Sitter Space,''
  Commun.\ Math.\ Phys.\  {\bf 87}, 577 (1983).
  %%CITATION = CMPHA,87,577;%%
  
  %\cite{Gubser:1996de}
\bibitem{Gubser:1996de}
  S.~S.~Gubser, I.~R.~Klebanov and A.~W.~Peet,
  ``Entropy and Temperature of Black 3-Branes,''
  Phys.\ Rev.\ D {\bf 54}, 3915 (1996)
  [arXiv:hep-th/9602135].
  %%CITATION = HEP-TH 9602135;%%
   
   %\cite{Constable:1999gb}
\bibitem{Constable:1999gb}
  N.~R.~Constable and R.~C.~Myers,
  ``Spin-two glueballs, positive energy theorems and the AdS/CFT
  correspondence,''
  JHEP {\bf 9910}, 037 (1999)
  [arXiv:hep-th/9908175].
  %%CITATION = HEP-TH 9908175;%%
 
   
  %\cite{Horowitz:2005vp}
\bibitem{Horowitz:2005vp}
  G.~T.~Horowitz,
  ``Tachyon condensation and black strings,''
  JHEP {\bf 0508}, 091 (2005)
  [arXiv:hep-th/0506166].
  %%CITATION = HEP-TH 0506166;%%


\end{thebibliography}
\end{document}